\documentclass[aps,prl,floatfix,twocolumn,showpacs]{revtex4}%
\usepackage{amssymb}
\usepackage{amsmath}
\usepackage{graphicx}
\usepackage{amsfonts}%
\providecommand{\U}[1]{\protect\rule{.1in}{.1in}}
\providecommand{\U}[1]{\protect\rule{.1in}{.1in}}
\providecommand{\U}[1]{\protect\rule{.1in}{.1in}}
\providecommand{\U}[1]{\protect\rule{.1in}{.1in}}
\providecommand{\U}[1]{\protect\rule{.1in}{.1in}}
\providecommand{\U}[1]{\protect\rule{.1in}{.1in}}
\providecommand{\U}[1]{\protect\rule{.1in}{.1in}}
\providecommand{\U}[1]{\protect\rule{.1in}{.1in}}
\begin{document}
\title{Inducing odd-frequency triplet superconducting correlations in a normal metal}
\author{Audrey Cottet}
\affiliation{Laboratoire Pierre Aigrain, Ecole Normale Sup\'{e}rieure, CNRS (UMR 8551),
Universit\'{e} P. et M. Curie, Universit\'{e} D. Diderot, 24 rue Lhomond,
75231 Paris Cedex 05, France}
\date{\today}

\pacs{73.23.-b, 74.45.+c, 85.75.-d}

\begin{abstract}
This work discusses theoretically the interplay between the superconducting
and ferromagnetic proximity effects, in a diffusive normal metal strip in
contact with a superconductor and a non-uniformly magnetized ferromagnetic
insulator. The quasiparticle density of states of the normal metal shows clear
qualitative signatures of triplet correlations with spin one (TCS1). When one
goes away from the superconducting contact, TCS1 focus at zero energy under
the form of a peak surrounded by dips, which show a typical spatial scaling
behavior. This effect can coexist with a focusing of singlet correlations and
triplet correlations with spin zero at finite but subgap energies. The
simultaneous observation of both effects would enable an unambigous
characterization of TCS1.

\end{abstract}
\maketitle

Hybrid superconducting/ferromagnetic circuits allow to obtain fascinating
unconventional superconducting correlations\cite{RevueBuzdin}. Although a
standard s-wave superconductor naturally hosts even-frequency superconducting
correlations, the superconducting proximity effect in an adjacent ferromagnet
can lead to odd-frequency pairing, because the ferromagnetic exchange field
lifts time reversal symmetry. This can correspond to either triplet
superconducting correlations with spin zero (TCS0), i.e. correlations between
opposite electronic spins, or triplet superconducting correlations with spin
one (TCS1), i.e. correlations between equal spins. The existence of TCS0 is
confirmed experimentally since a decade (see e.g.
\cite{KontosDOS,Ryazanov,KontosI}). The possibility of obtaining TCS1 has been
investigated more recently\cite{RevueBergeret}. The features observed so far
are quantitative. It has been observed that, in certain conditions,
ferromagnets can sustain a supercurrent on a much longer lengthscale than
expected\cite{Keizer}. This suggests the presence of TCS1, because in
diffusive ferromagnets, TCS1 are expected to propagate on a much longer
distance than TCS0\cite{Eschrig}. Most of the strategies discussed so far to
observe TCS1 require to measure a supercurrent, which is an energy-integrated
quantity. Alternatively, the quasiparticle density of states (DOS) yields
spectroscopic information and thus appears as a powerful tool to characterize
the nature of superconducting correlations. This paper presents a geometry in
which the DOS gives a particularly rich access to odd-frequency
superconducting correlations.\begin{figure}[ptb]
{\small \centering\includegraphics[width=0.6\linewidth]{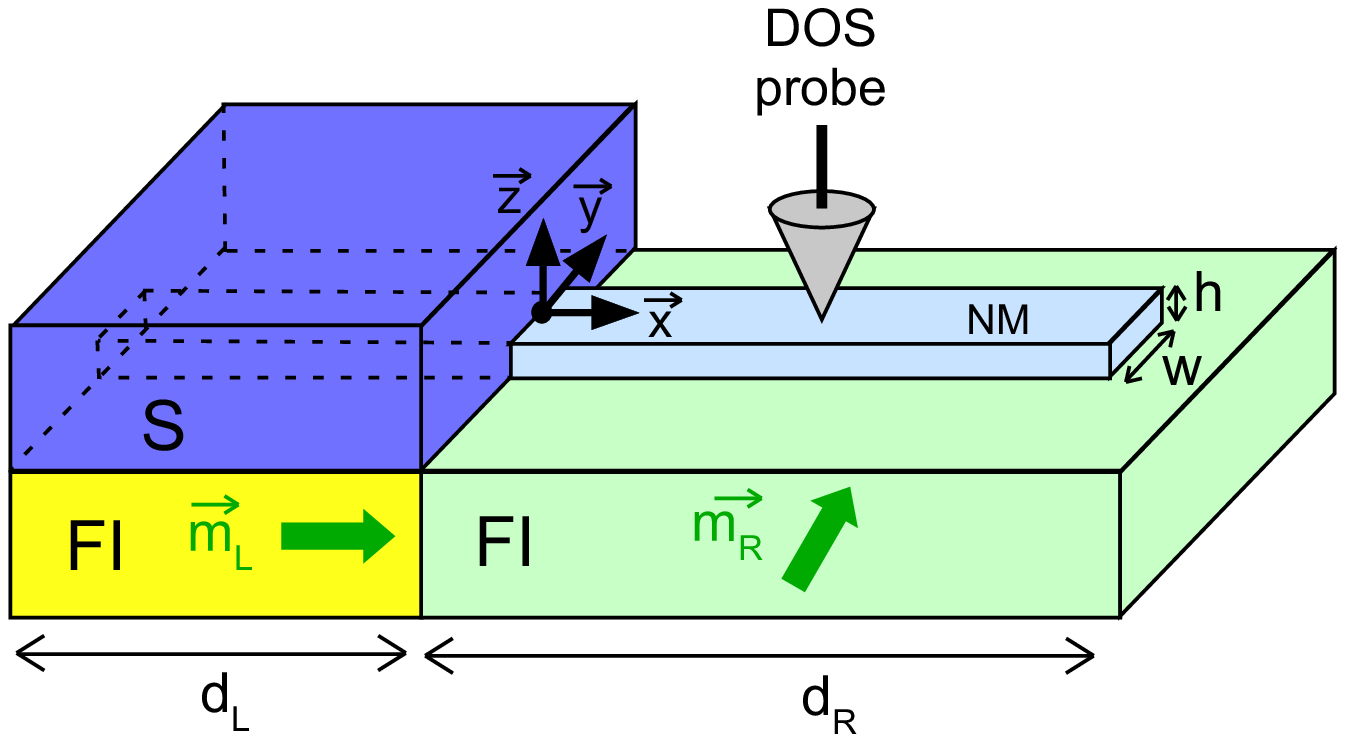}\newline%
}\caption{Scheme of the lateral geometry considered in this work (see text).}%
\end{figure}

I suggest to use a lateral geometry, where the superconducting correlations
are induced in a normal metal (NM) strip in contact with a superconductor (S)
and a ferromagnetic insulator (FI) with two non-colinear magnetization
domains. This configuration is compatible with spatially resolved DOS
measurements using several tunnel contact probes\cite{Gueron} or a low
temperature STM\cite{Le Sueur}. The propagation of odd-frequency
superconducting correlations has been studied thoroughly in
ferromagnets\cite{RevueBergeret}, but only elusively in
NMs\cite{Asano,Yokoyama,Yokoyama3}. Here, the FI turns the NM strip into an
effective ferromagnet with an unusually weak exchange field. This is
advantageous to study the propagation of odd-frequency superconducting
correlations. The TCS1 induce a zero-energy peak in the DOS of the
NM\cite{Asano}. Such an effect is not specific to
TCS1\cite{Fauchere,Krawiec,Linder2,Yokoyama2}. However, in the present
geometry, important additional features allow to identity unambiguously TCS1.
Indeed, the low-energy DOS peak is surrounded by dips, and this structure
shows a characteristic spatial scaling behavior when one goes away from the
superconducting contact, i.e. it "shrinks" on a lenghtscale which depends on a
longitudinal Thouless energy. If one observes simultaneously finite energy
dips, which confirm the existence of the effective exchange field inside the
NM, the zero-energy peak points unambiguously to TCS1.

I consider the lateral geometry of Fig.1, where the central element is a NM
strip with thickness $h$, width $w$, and longitudinal coordinate $x$. A
portion with length $d_{L}$ of the NM strip is contacted to a S and a FI
domain magnetized along a direction $\vec{m}_{L}$ for $x<0$, and a portion
with length $d_{R}$ is contacted to a FI domain magnetized along a direction
$\vec{m}_{R}$ for $x>0$. I assume that the structure is diffusive, so that a
quasiclassical isotropic Green's function $G$ can be used to describe the
propagation of superconducting correlations inside the NM\cite{conventions}.
The function $G$ has a structure in the spin and Nambu (electron-hole)
subspaces, so that it can be decomposed in terms of the spin[Nambu] Pauli
matrices $\sigma_{i}$[$\tau_{i}$] (see below). In the following, I use
$\vec{m}_{L}=\vec{x}$. In the colinear case $\vec{m}_{R}=\vec{x}$, only
singlet correlations and TCS0 can appear inside the NM. In the non-colinear
case $\vec{m}_{R}\neq\vec{x}$, TCS1 can also appear.

The spatial evolution of $G$ inside the NM is described by the Usadel equation
$\hbar D\rho\vec{\nabla}\mathbf{.}\vec{j}=-(i\varepsilon-\Gamma)\left[
\check{\tau}_{3},G\right]  $, with $\rho$ the resistivity of the
NM\cite{Usadel}. The density of matrix current $\vec{j}=G\vec{\nabla}G/\rho$
characterizes the flows of charge, spin and electron-hole coherence in the
device. The rate $\Gamma$ accounts for inelastic processes. The Usadel
equation alone is not sufficient to predict the behavior of $G$, because the
influence of the $S$ and $FI$ contacts must be taken into account. When $h$
and $w$ are smaller than the typical spatial scale $\xi$ characterizing the
variations of $G$ inside the NM\cite{xi}, it is possible to derive a
one-dimensional effective Usadel equation on $G\simeq G(x)$, which takes into
account the effects of the S and FI contacts. This equation can be derived
with an approach inspired from circuit theory\cite{CircuitTheory}, using spin
dependent boundary conditions for isotropic Green's functions\cite{BCIGF}.
This requires to define a surface tunnel conductance $G_{T}^{s}$ for the S/NM
interface and a surface conductance-like coefficient $G_{\phi}^{s}$ for the
NM/FI interface. The last parameter accounts for the fact that electrons from
the NM are reflected by the FI with spin-dependent reflection phases. Indeed,
the internal Stoner exchange field of the FI can affect the evanescent tails
of electronic wavefunctions on a scale of a few atomic layers. One finds, for
$x<0$
\begin{equation}
2l^{2}\mathbf{\nabla}_{x}(G\mathbf{\nabla}_{x}G)=-2\gamma_{\varepsilon}\left[
\check{\tau}_{3},G\right]  -\gamma_{T}\left[  G,G_{S}\right]  +i\gamma_{\phi
}[\tau_{3}\sigma_{L},G] \label{UsadelEff}%
\end{equation}
with $E_{Th}=\hbar D/h^{2}$, $\gamma_{\varepsilon}=(i\varepsilon
-\Gamma)/E_{Th},\gamma_{T}=G_{T}^{s}\rho h$, $\gamma_{\phi}=G_{\phi}^{s}\rho
h$ and $\sigma_{L(R)}=\vec{m}_{L(R)}.\vec{\sigma}$. Here, $G_{S}$ denotes the
value of the isotropic Green's function inside the superconductor. For $x>0$,
one finds a similar equation with $\gamma_{T}$ replaced by $0$ and $\sigma
_{L}$ by $\sigma_{R}$.\begin{figure}[ptb]
{\small \centering\includegraphics[width=0.75\linewidth]{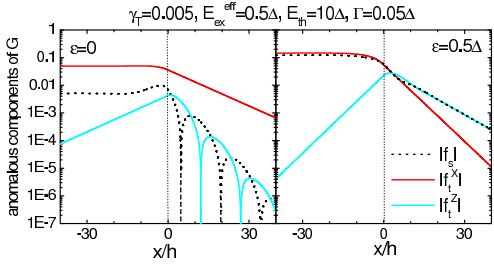}
}\caption{Absolute values of the anomalous components $f_{s}$, $f_{t}^{x}$ and
$f_{t}^{z}$ of $G$ in the NM strip at $\epsilon=0$ (left panel) and
$\epsilon=E_{ex}^{eff}$ (right panel) respectively, for $\vec{m}_{R}=\vec{z}$
and $d_{L(R)}\longrightarrow+\infty$. }%
\end{figure}

Equation (\ref{UsadelEff}) describes an interplay between the superconducting
and ferromagnetic proximity effects. On the one hand, the $\gamma_{T}$ terms
tend to induce a minigap inside the NM, due to the confinement of electrons in
the $\vec{z}$ direction. This effect depends on the interface parameter
$\gamma_{T}\ $but also the lateral Thouless energy $E_{Th}$ defined below
Eq.~(\ref{UsadelEff}). For instance, at the left end of the NM strip, the DOS
is suppressed for energies $\left\vert \varepsilon\right\vert <\tilde{\Delta
}=E_{Th}\gamma_{T}/2$ when $d_{L}\rightarrow+\infty$, $E_{Th}\gamma_{T}%
\ll\Delta$ and $\gamma_{\phi}=0$. On the other hand, the interface parameter
$G_{\phi}^{s}$ causes an effective exchange field $E_{ex}^{eff}=E_{Th}%
\gamma_{\phi}/2=\hbar DG_{\phi}^{s}\rho/h$ oriented along $\vec{x}$ [$\vec
{m}_{R}$] at the left[right] side of the NM strip, i.e. $x<0$ [$x>0$]
\cite{Huertas-Hernando}. This type of effective field has already been
observed and accurately characterized for various types of S/FI
bilayers\cite{Tedrow,Hao,Meservey,Xiong}.\ For $x<0$, the exchange field due
to the left contact splits the minigap induced by the S contact and induces
TCS0 with respect to the $\vec{x}$ direction. These correlations can propagate
to the right part of the NM strip where they correspond to TCS1 when $\vec
{m}_{R}=\vec{z}$. The induction of TCS1 in diffusive S/F structures present
similarities with this process\cite{RevueBergeret}. However, minigap effects
are usually destroyed in ferromagnets, due to strong ferromagnetic exchange
fields, except in some particular disordered cases\cite{Ivanov}. Besides,
$E_{ex}^{eff}$ is expected to be much smaller than the exchange field inside
standard ferromagnets. Hence, the situation studied here is qualitatively
different from that of standard S/F structures. Interestingly, for a given
type of NM/FI contact, the amplitude of $E_{ex}^{eff}$ can be tuned by
choosing $h$ at the sample fabrication stage, since $E_{ex}^{eff}$ scales with
$1/h$ \cite{Hao,Huertas-Hernando,Cottet}. This gives an interesting
flexibility with respect to material constraints.

In the following, I assume that the value $G_{S}$ of $G$ inside the
superconductor is equal to the bulk BCS value, i.e. $G_{S}=\cos(\theta
_{S})\check{\tau}_{3}+\sin(\theta_{S})\check{\tau}_{1}$, with $\theta
_{S}=\arctan[\Delta/(-i\varepsilon+\Gamma)]$. Inside the NM, one can use the
angular parametrization $G=\tau_{3}(\cosh\eta)\cos(\theta)+i\sinh(\eta
)\sin(\theta)\vec{v}.\vec{\sigma})$+$\tau_{1}(f_{s}+\vec{f}_{t}.\vec{\sigma}%
)$, with $f_{s}=\cosh(\eta)\sin(\theta)$, $\vec{f}_{t}=-i\sinh(\eta
)\cos(\theta)\vec{v}$, and $\vec{v}$ a unit vector. This convention
automatically satisfies the normalization condition $G^{2}=1$. I discuss below
the anomalous components of $G$, i.e. $f_{s}$, $f_{t}^{x}=\vec{f}_{t}.\vec{x}$
and $f_{t}^{z}=\vec{f}_{t}.\vec{z}$, which reveal the existence of
superconducting correlations inside the NM. Defining TCS0 and TCS1 components
requires to define a reference direction $\vec{m}_{ref}$. A natural choice is
to use $\vec{m}_{ref}=\vec{m}_{L}$ for $x<0$ and $\vec{m}_{ref}=\vec{m}_{R}$
for $x>0$. Then, TCS0 and TCS1 correspond to the components of $\vec{f}_{t}$
parallel and perpendicular to $\vec{m}_{ref}$, respectively. The DOS inside
the strip can be calculated as $N(\varepsilon)/N_{0}=%
{\textstyle\sum\nolimits_{\sigma}}
\operatorname{Re}[\cos(\theta+i\sigma\eta)]/2$, with $N_{0}$ the normal state
DOS of the NM.

The spatial evolution of the DOS can be first studied with an analytic
linearized description, which yields a transparent interpretation of the
circuit behavior. This approach is valid in the limit of a weak
superconducting proximity effect, i.e. $\theta,\eta\ll1$ inside the NM. This
occurs e.g. when $\gamma_{T}$ is small and $\Gamma$ large, so that the minigap
is closed and just gives residual dips at the left side of the NM strip. For
simplicity, I assume $d_{L(R)}\longrightarrow+\infty$. For $x<0$ one has
\begin{equation}
\lbrack f_{s},f_{t}^{x},f_{t}^{z}]=[0,0,1]A_{0}e^{k_{S}x}+%
{\displaystyle\sum\limits_{\sigma=\pm1}}
[1,\sigma,0]\left(  \frac{\Theta_{\sigma}^{B}}{2}+A_{\sigma}e^{-\tilde
{k}_{\sigma}x}\right)  \label{2}%
\end{equation}
with $k_{S}^{2}h^{2}-\gamma_{T}\cos(\theta_{S})=-2i[(\varepsilon
+i\Gamma)/E_{Th}]$, and $\tilde{k}_{\sigma}h^{2}-\gamma_{T}\cos(\theta
_{S})=2i[\sigma E_{ex}^{eff}-\varepsilon-i\Gamma]/E_{Th}$. Here, I note
\begin{equation}
\Theta_{\sigma}^{B}=\arctan[\frac{\gamma_{T}\sin\left[  \theta_{S}\right]
}{\gamma_{T}\cos\left[  \theta_{S}\right]  +2[-i\epsilon+\Gamma+i\sigma
E_{ex}^{eff}]/E_{Th}}] \label{3}%
\end{equation}
the value of $\theta-i\eta\sigma$ at $x\rightarrow-\infty$. {}For $x>0$ and
$\vec{m}_{R}=\vec{z}$, one has
\begin{equation}
\lbrack f_{s},f_{t}^{x},f_{t}^{z}]=[0,1,0]B_{0}e^{-k_{0}x}+%
{\displaystyle\sum\limits_{\sigma=\pm1}}
[1,0,\sigma]B_{\sigma}e^{-k_{\sigma}x} \label{Lin}%
\end{equation}
with $k_{0}^{2}h^{2}=-2i[(\varepsilon+i\Gamma)/E_{Th}]$ and $k_{\sigma}%
h^{2}=2i[\sigma E_{ex}^{eff}-\varepsilon-i\Gamma]/E_{Th}$. For $x>0$ and
$\vec{m}_{R}=\vec{x}$, the second and third components of the vectors in
Eq.(\ref{Lin}) must be exchanged. The coefficients $A_{\sigma}$, $A_{0}$,
$B_{\sigma}$, and $B_{0}$ can be calculated by assuming the continuity of
$f_{s},f_{t}^{x},f_{t}^{z}$ and their first derivatives at $x=0$. This leads
to the results shown in Fig.~2, obtained for $\vec{m}_{R}=\vec{z}$
(non-colinear case). \begin{figure}[ptb]
{\small \centering\includegraphics[width=0.7\linewidth]{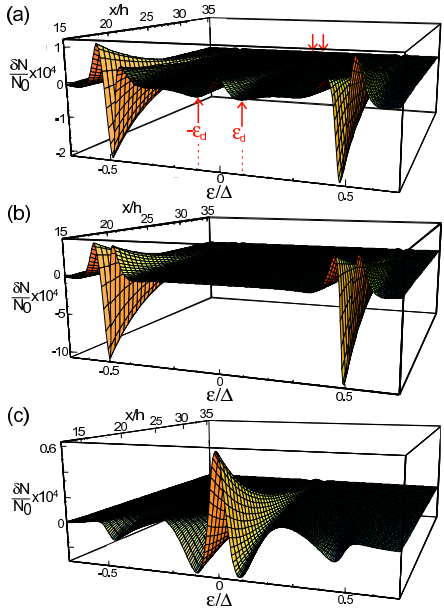}\newline%
}\caption{Density of states in the NM in the non-colinear case $\vec{m}%
_{R}=\vec{z}$ (panel a), the colinear case $\vec{m}_{R}=\vec{x}$ (panel b),
and the case where there is no FI contact at $x>0$ (panel c). The data are
shown in terms of the deviation $\delta N=N-N_{0}$ of the DOS in the NM from
the normal state value $N_{0}$. The parameters used are the same as in
Fig.~2.}%
\end{figure}At zero-energy and $x\rightarrow-\infty$, $f_{t}^{x}$ is dominant
due to $\gamma_{T}\ll\gamma_{\phi}$ (Fig.~2, left panel)\cite{Linder1}. I now
comment the behavior of the different types of correlations for $x>0$ (Fig.~2,
right panel). In this area, TCS1 ($f_{t}^{x}$) propagate independently of the
two other components, with the characteristic vector $k_{0}$. At
$\varepsilon=0$, $k_{0}$ is small, so that TCS1 propagate on a relatively long
distance: this is why TCS1 are usually called "long-range" correlations.
However, this is true only at low energies. Indeed, for $\varepsilon=\pm
E_{ex}^{eff}$, \ the real part of $k_{0}$ is larger, so that TCS1 decay more
quickly. Conversely, for $x>0$ and $\varepsilon=0$, the propagation of singlet
correlations ($f_{s}$) and TCS0 ($f_{t}^{z}$) is short-range. Indeed, $f_{s}$
and $f_{t}^{z}$ show a damped oscillatory behavior ruled by the propagation
vectors $k_{\sigma}$, which correspond to a scale $(\hbar D/E_{ex}%
^{eff})^{1/2}$ for $E_{ex}^{eff}\gg\Gamma$ and $\varepsilon=0$. In contrast,
for $\varepsilon=\pm E_{ex}^{eff}$, the $f_{s}$ and $f_{t}^{z}$ components do
not oscillate and decay more slowly. Hence, it is the singlet correlations and
TCS0 which are long-range for $\varepsilon=\pm E_{ex}^{eff}$. As a result,
sufficiently far from the superconductor, TCS1 become dominant at
$\varepsilon=0$, whereas singlet correlations and TCS0 become dominant for
\ $\varepsilon=\pm E_{ex}^{eff}$. In other terms, one obtains $\vec
{v}(\varepsilon=0)\simeq\vec{x}$ and $\vec{v}(\varepsilon=\pm E_{ex}%
^{eff})\simeq\vec{z}$. This leads to characteristic features in the energy
dependence of the DOS, shown by Fig.~3a. For $x$ sufficiently large, one
obtains characteristic dips at $\varepsilon\simeq\pm E_{ex}^{eff}$, due to
singlet correlations and TCS0. In contrast, TCS1 produce a zero-energy peak
which reaches a maximum higher than the normal-state DOS $N_{0}$. Importantly,
$N(0)>N_{0}$ is not specific to TCS1. Indeed, an enhanced zero-energy DOS can
also be due for instance to interaction effects\cite{Fauchere}, or to
TCS0\cite{Buzdin,Linder1,Linder2,Yokoyama2}, as observed experimentally in S/F
bilayers\cite{KontosDOS}. However, in the present geometry, TCS1 can be
detected unambiguously due to the additional features discussed below.

Due to the peculiar energy dependence of $k_{0}$, the zero-energy DOS peak is
surrounded by low-energy dips, and this ensemble shows a characteristic
spatial scaling behavior when $x$ increases (see Fig. 3a). Let us note
$\pm\varepsilon_{d}$ the position of the low-energy dips. In the limit of a
vanishing $\Gamma$, $\varepsilon_{d}$ scales with a longitudinal Thouless
energy $\tilde{E}_{Th}=\hbar D/x^{2}$, provided the finite energy dips are
well separated from the low-energy peak, i.e. $\varepsilon_{d}\ll E_{ex}%
^{eff}$. When $\Gamma$ is finite, a scaling behavior can persist. For
instance, in the limit $\varepsilon_{d}\ll\Gamma,E_{ex}^{eff}$, Eq.~(\ref{Lin}%
) gives
\begin{equation}
\varepsilon_{d}\simeq\pi\sqrt{\Gamma\tilde{E}_{Th}/2}\varpropto1/x \label{lim}%
\end{equation}
The scaling behavior of the low energy DOS features represents a "smoking-gun"
for the fact that some superconducting correlations propagate along the strip
with the vector $k_{0}$. Importantly, the DOS dips at $\varepsilon\simeq\pm
E_{ex}^{eff}$ follow an equivalent scaling behavior, due to the structure of
the $k_{\sigma}$ vectors. It is instructive to discuss other magnetic
configurations. Figure 3.b shows the DOS of the NM in the colinear case. The
finite energy dips are still present, but there is no low-energy features
because TCS1 are absent. At last, Fig 3c. shows the DOS of the NM when there
is a FI contact with a uniform magnetization at $x<0$, but no FI contact or
$E_{ex}^{eff}=0$ at $x>0$. In this case, for $x>0$, weak DOS dips appear at
$\varepsilon\simeq\pm E_{ex}^{eff}$, due to the spin-split minigap effect
occurring for $x<0$. Importantly, these dips are very different from those
obtained in the previous cases: they are not surrounded by peaks and quickly
vanish with increasing $x$. In contrast, there is still a spatially-scaling
zero-energy peak, although there is no TCS1 in the circuit. This is because
singlet correlations and TCS0 propagate with the characteristic vector $k_{0}$
in this case. One can conclude that it is important to see DOS dips appear at
$\varepsilon\simeq\pm E_{ex}^{eff}$ and persist with increasing $x$, to
confirm the existence of $E_{ex}^{eff}$ at the right side of the strip. In
this case, the spatially-scaling zero-energy peak can only be due to TCS1,
which is the only component which can propagate with $k_{0}$. For a good
visibility of the finite-energy DOS dips, it is important to use $E_{ex}%
^{eff}<\Delta$. In practice, this limit can be reached by using an appropriate
value for $h$. Note that the long-range behavior of TCS0 and singlet
correlations at $\varepsilon=\pm E_{ex}^{eff}$ is usually not discussed for
standard ferromagnets, in which exchange fields are too high.

\begin{figure}[ptb]
{\small \centering\includegraphics[width=0.6\linewidth]{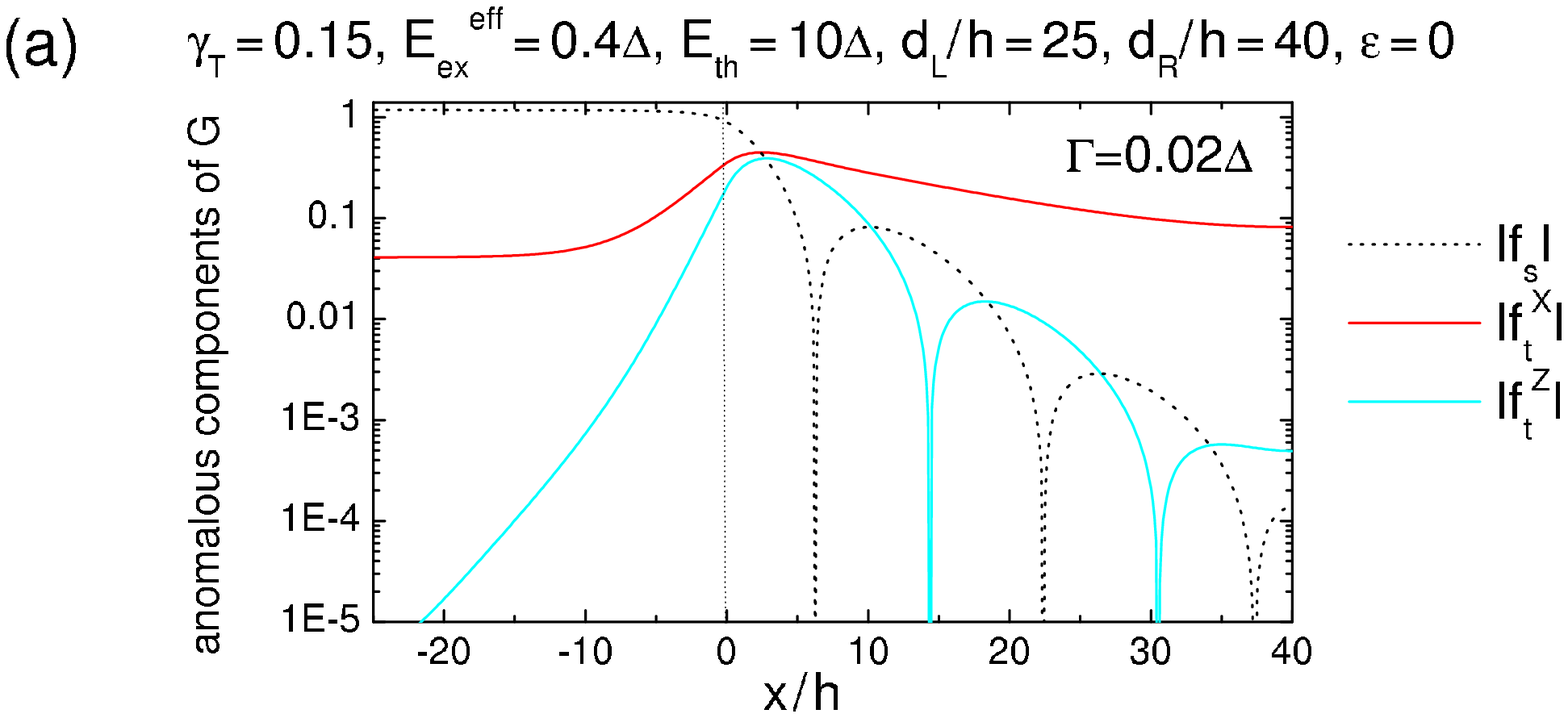}\newline%
\newline\includegraphics[width=0.4\linewidth]{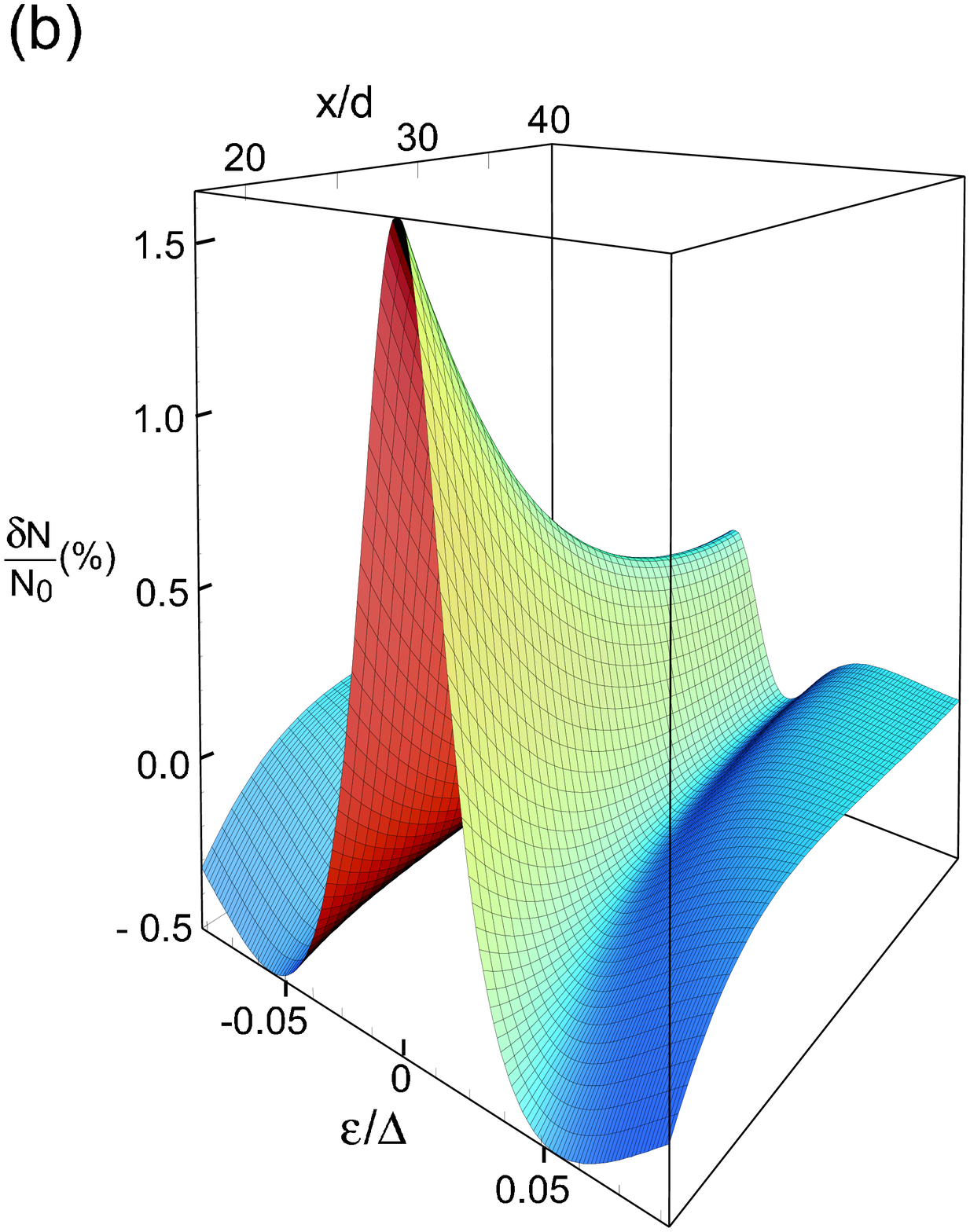}\quad
\includegraphics[width=0.3\linewidth]{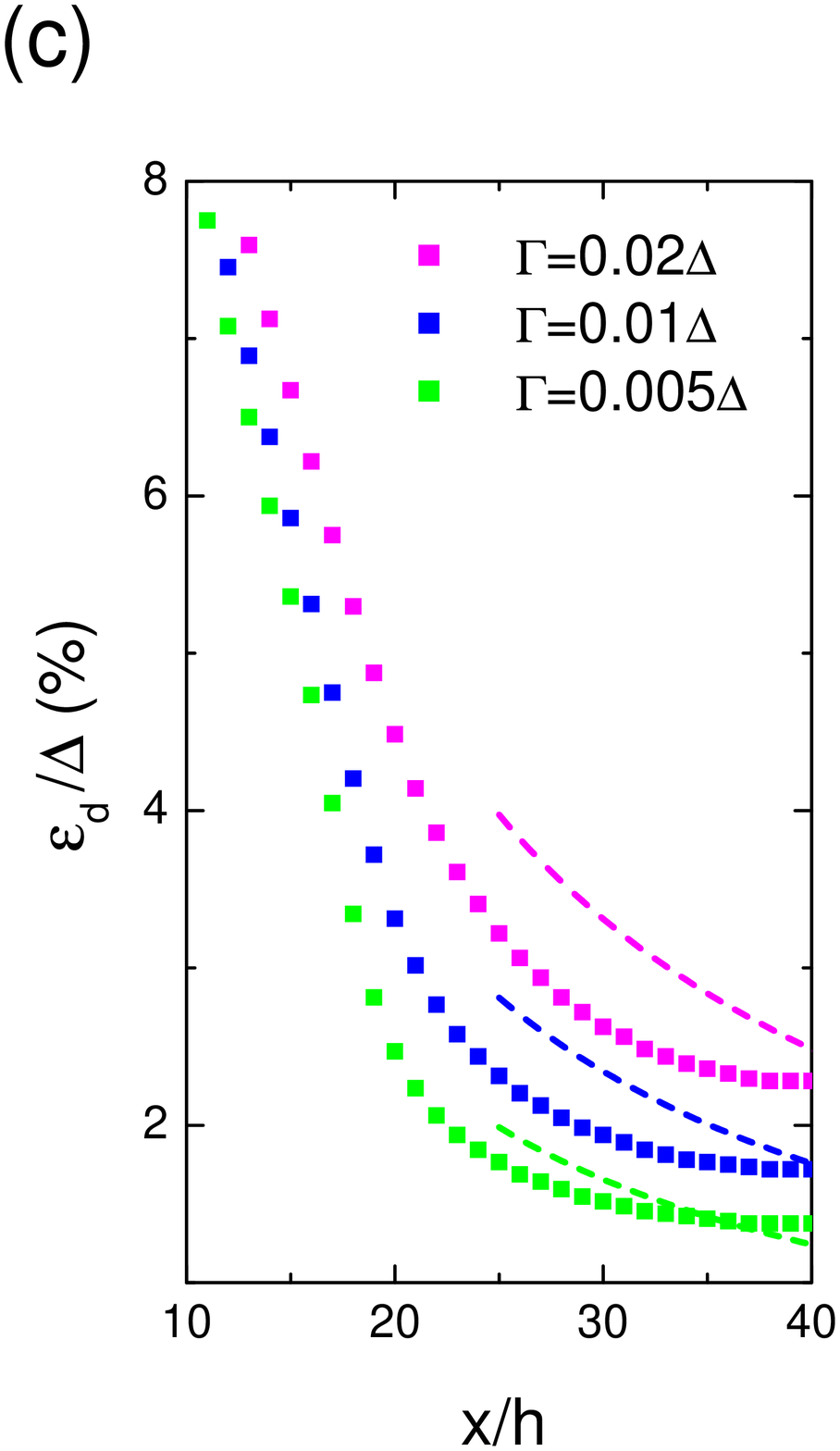}\newline}%
\caption{Predictions for $\vec{m}_{R}=\vec{z}$, a value of $\gamma_{T}$ much
larger than in Figs. 2 and 3, and finite values of $d_{L(R)}$. Panel a shows
$\left\vert f_{s}\right\vert $, $\left\vert f_{t}^{x}\right\vert $ and
$\left\vert f_{t}^{z}\right\vert $ at zero energy, as a function of the
coordinate $x$. Panel b shows $\delta N$ as a function of $\varepsilon$ and
$x$. Panel c shows the dependence of $\varepsilon_{d}$ on $x$, for various
values of $\Gamma$. The value $\Gamma=0.02\Delta$ is used in panels a and b.
All curves have been obtained with the numerical approach, except the dotted
lines in panel c, which correspond to Eq.(\ref{lim}).}%
\end{figure}

The amplitude of the DOS variations in Figs.~2 and 3 is very weak due to the
small $\gamma_{T}$ used. Experimental observations call for an increase in
$\gamma_{T}$. This requires to use a numerical approach to solve the problem
in the non-linear limit. In practice, ferromagnetic domains have a finite
size, so that $d_{L}$ and $d_{R}$ must be finite. The resulting finite size
effects are studied below with the numerical approach. Figure 4 presents
results for the non-colinear case. The spatial behaviors of $f_{t}^{x}$,
$f_{t}^{z}$ and $f_{s}$ remain qualitatively similar (see Fig 4.a). At $x>0$,
$f_{t}^{x}$ still yields the expected low-energy features in the DOS (Fig.4b).
The variations of $\varepsilon_{d}$ with $x$ are shown in Fig.4c for various
values of $\Gamma$ (symbols). The scaling behavior of the DOS low-energy
features is robust to finite size effects. One can check that the
semi-infinite expression Eq. (\ref{lim}) yields the right order of magnitude
for $\varepsilon_{d}$ at the right boundary of the NM strip. Finite size
effects slow down the scaling behavior of the low-energy features at
$x\longrightarrow d_{R}$, i.e. one has $\partial\varepsilon_{d}/\partial
x\longrightarrow0$. Nevertheless, it is still possible to observe a strong
decrease of $\varepsilon_{d}$ with $x$, when $x$ is not too close to $d_{R}$.
If $\gamma_{T}$ is too large with respect to $\gamma_{\phi}$, the zero-energy
DOS peak has a reduced amplitude because $f_{t}^{x}$ is weak\cite{Linder1}. To
maximize $N(0)$, one must use $\gamma_{T}=\gamma_{\phi}$ and decrease $\Gamma
$. For instance, using $\gamma_{T}=0.08$, $\Gamma=0.005\Delta$ and the other
parameters of Fig.4.b (in particular, $\gamma_{\phi}=0.08$), one obtains
$N(0)=1.35N_{0}$ at $x=40h$. Nevertheless, using unfavorable parameters like
those of Fig.4.b, one still obtains $N(0)=1.00325N_{0}$ at $x=40h$, and
amplitude which is measurable, in principle\cite{KontosDOS}. The geometry
discussed here also presents the advantage that spin-flip scattering effects,
which could reduce the amplitude of odd-frequency superconducting
correlations, are usually small in NMs. Interestingly, a NM strip with a S
contact but no ferromagnetic contacts has been studied experimentally
($E_{ex}^{eff}=0$ for any $x$). In this case, one can only have singlet
correlations propagating with $k_{0}$. As a result, a spatially-scaling
zero-energy dip has been observed in the DOS of the NM\cite{Gueron,BelzigTM}.

To conclude, TCS1 can appear in a diffusive NM strip in contact with a S and a
FI with several non-colinear magnetic domains. These correlations induce in
the DOS of the NM a low-energy peak surrounded by dips, which show a
characteristic spatial scaling behavior away from the S contact. Meanwhile, if
the thickness of the strip is chosen properly, superconducting correlations
between opposite spins will focus at finite but subgap energies. The
simultaneous observation of both effects would enable an unambiguous
identification of TCS1.

\textit{I thank T. Kontos and W. Belzig for useful discussions.}


\begin{thebibliography}{99}
\bibitem {conventions}The function $G$ used here is related to the function
$G^{\prime}$ defined in Ref.\cite{BCIGF} by $G=TG^{\prime}T$, with
$T=[(1+\tau_{3})\sigma_{3}+(1-\tau_{3})\sigma_{2}]/2$.
Refs.~\cite{Ivanov,Yokoyama} use a similar convention.
\bibitem {xi}The various components of $G$ can propagate on different spatial
scales. For simplicity, I note $\xi$ the minimum of these scales throughout
the structure.
\end{thebibliography}
\end{document}